\documentclass[showpacs,twocolumn,aps,pra]{revtex4}
\usepackage{amssymb}
\usepackage{amsmath}
\usepackage{graphicx}
\usepackage{epsfig}

\begin{document}

\title{Rigorous results for tight-binding networks: particle trapping and
scattering}
\author{L. Jin and Z. Song}
\email{songtc@nankai.edu.cn}
\affiliation{School of Physics, Nankai University, Tianjin 300071, China}

\begin{abstract}
We investigate the particle trapping and scattering properties in a
tight-binding network which consists of several subgraphs. The particle
trapping condition is proved under which particles can be trapped in a
subgraph without leaking. Based on exact solutions for the configuration of
a $\pi$-shaped lattice, it is argued that the bound states in a specified
subgraph are of two types, resonant and evanescent. We also link the
trapping rigorous result to the scattering problem. The scattering features
of the $\pi $-shaped lattice is investigated in the framework of the Bethe
Ansatz.
\end{abstract}

\maketitle


\section{Introduction}

Trapping and scattering of a particle is an important feature in many
quantum information processing systems. Due to the development of
technology, the implementation of quantum information processing in quantum
systems with periodic potential, such as optical lattices \cite{OL}, arrays
of quantum dots \cite{QD}, photonic crystal \cite{PC} and coupled-resonator
optical waveguide \cite{CROW}, has attracted intensive investigations. The
design of quantum device based on these promising technologies relies on the
particle trapping and scattering properties in a discrete system. A
heuristic example shows that the quantum confinement in a discrete system is
distinct from its counterpart in continuum media \cite{Longhi PRE}, due to
the Wannier-Stark localization \cite{Fukuyama}.

This paper focuses on noninteracting particles on discrete lattice, which is
treated by tight-binding approximation. Intuitively, the particle trapping
is implemented by sufficient strong on-site potential as the continuous
system. In contrast to continuum, however, different dynamical properties
emerge in the lattice system due to its distinct dispersion relation: a
local wave packet can be confined by linear potential distribution \cite%
{Longhi PRE} and the degree of spreading of a propagating wave packet can be
controlled by judicious choice of the particle energy \cite{Osborne, Yang
PRA, Kim PRB}. Recent studies show that Fano resonance may be employed to
construct the perfect mirror or transparency so as to control particles in a
region of the lattice \cite{ZhouL1, ZhouL2, Liao} via engineered
configurations. Because of the numerous varieties of the possible geometry
of the quantum network, we believe it is beneficial to have lattice-based
rigorous results and exact solutions for the devise of a quantum device. In
this paper, we show rigorously that the perfect particle trapping without
any leakage can be achieved in simple tight-binding networks. This provides
a method to devise the quantum network to confine particles with required
mode. We also link the trapping rigorous result to the scattering problem.
This general finding is illustrated by a practical network consisting of a
waveguide with an embedded $\pi $-shaped subgraph. Exact solutions for such
types of configurations are obtained to demonstrate and supplement the
rigorous results.

\section{Rigorous result for particle trapping}

A general tight-binding network is constructed topologically by the sites
and the various connections between them, and is also represented as a
vertex-edge graph. Cutting off some of the connections, a graph is
decomposed into several subgraphs. So when a particle is strictly trapped
within a certain region of a network, one can say that it is confined in a
specified subgraph. The main aim of this paper is to answer the questions of
what kind of subgraph can trap a particle as bound state and of how such a
subgraph scatters a particle when it is embedded in a waveguide.
\begin{figure}[tbp]
\includegraphics[ bb=37 221 583 739, width=4.0cm, clip]{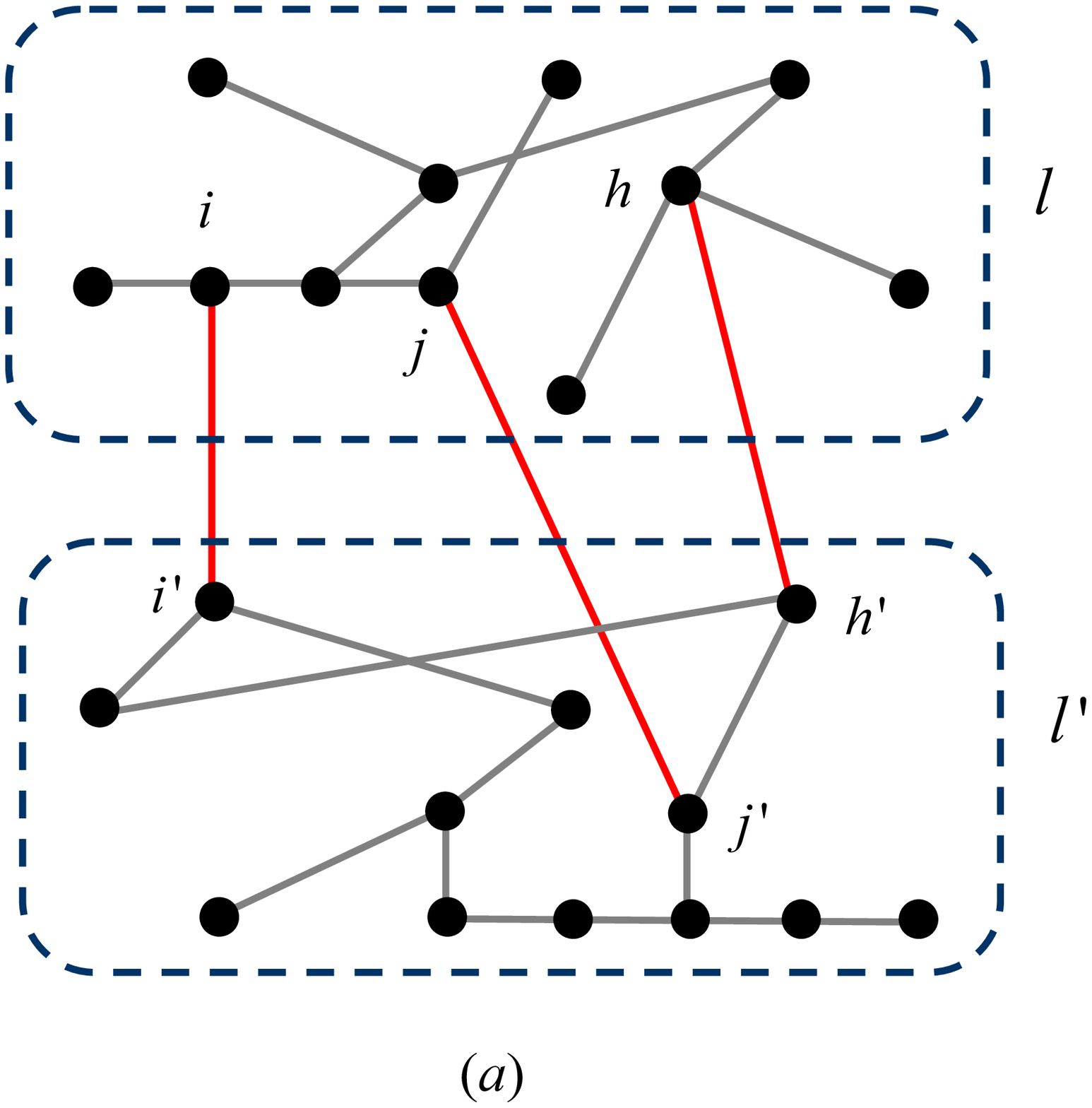}%
\includegraphics[ bb=37 221 583 739, width=4.0cm, clip]{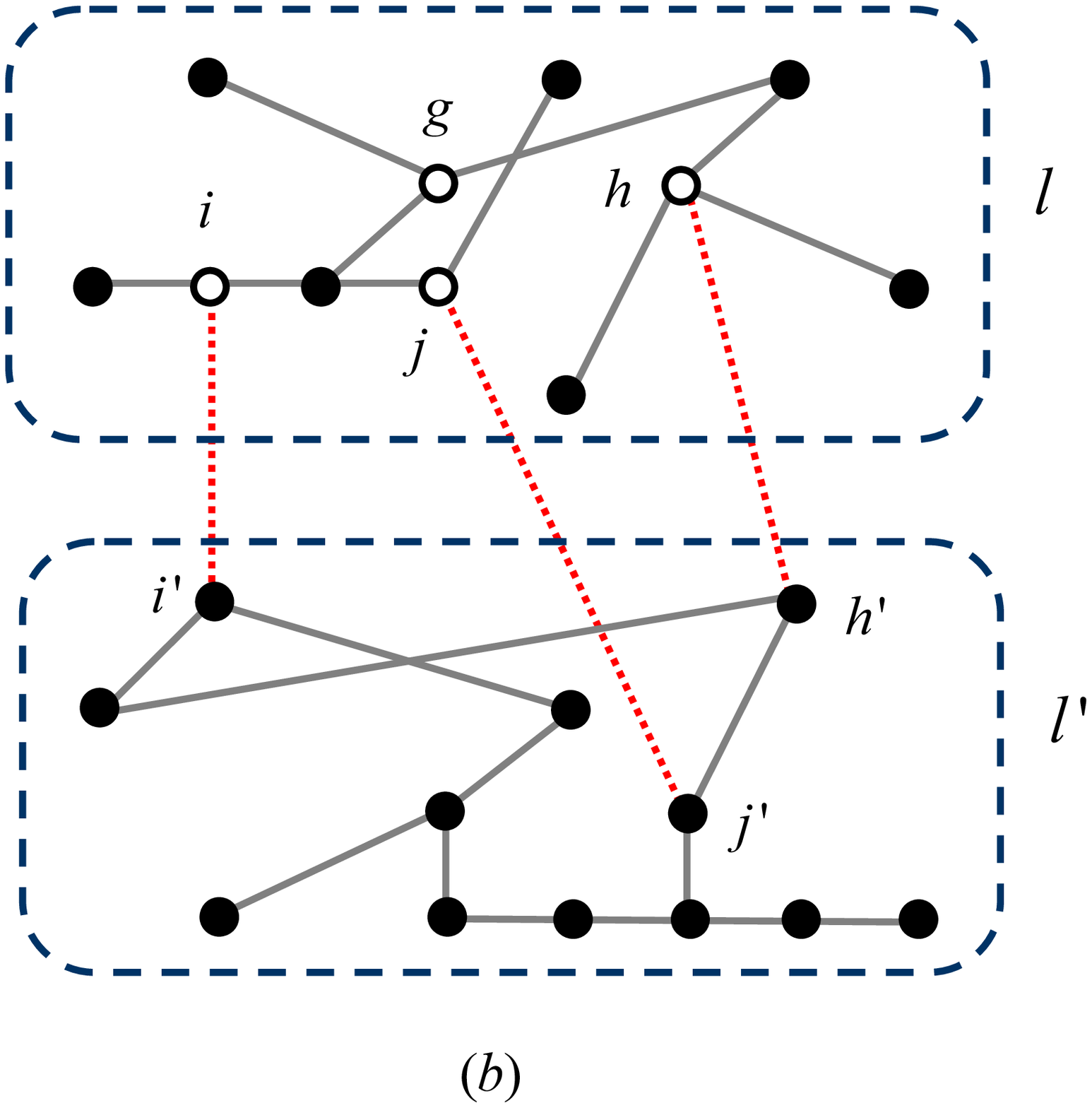}
\caption{(Color online) An arbitrary graph of tight-binding network within
part of which particles may be confined without any leakage. The graph can
be decomposed into two subgraphs $l$ and $l^{\prime }$ which are connected
via the coupling between the joint sites $(i,j,h)$ and $(i^{\prime
},j^{\prime },h^{\prime })$ (a). The perfect bound states can be formed in
subgraph $l$ when the eigen functions of $H_{l}$\ have wave nodes on all the
joint sites $(i,j,h)$, which are denoted by empty circles (b). The existence
of more wave nodes\ (like site $g$) may allow multiple bound states.}
\label{fig1}
\end{figure}
The Hamiltonian of a tight-binding network, or a graph which consists of $%
n_{0}$ subgraphs reads as
\begin{eqnarray}
H &=&\sum_{l=1}^{n_{0}}H_{l}+\sum_{lm}H_{lm},  \label{H} \\
H_{l} &=&-\sum_{\left\langle ij\right\rangle }(\kappa
_{ij}^{[l]}a_{l,i}^{\dag }a_{l,j}+\text{H.c.})+\sum_{i=1}^{N_{l}}\mu
_{i}^{[l]}a_{l,i}^{\dag }a_{l,i},  \notag \\
H_{lm} &=&-\sum_{i,j}(\kappa _{ij}^{[lm]}a_{l,i}^{\dag }a_{m,j}+\text{H.c.}),
\notag
\end{eqnarray}%
where label $l$ denotes the $l$th subgraph of $N_{l}$ site, which subgraph
is defined by the distribution of the hopping integrals $\{\kappa
_{ij}^{[l]}\}$ and on-site potentials $\{\mu _{i}^{[l]}\}$, and $%
a_{l,j}^{\dag }$ is the boson or fermion creation operator at the $j$th site
in the $l$th subgraph. Here, $H_{l}$ and $H_{lm}$ represent the Hamiltonians
of the subgraphs and the couplings between them. In terms $H_{lm}$, site $i$
$(j)$ is the \textit{joint site} of subgraph $l$ $(m)$ for the connections
to other subgraphs. Obviously, the decomposition of subgraphs is arbitrary,
and can be implemented at will. Figure \ref{fig1} shows an example
schematically. Note that the Hamiltonians $H_{l}$ (also $H$) are quadratic
in particle operators and can be diagonalized through the linear
transformation%
\begin{equation}
\eta _{l,k}^{\dag }=\sum_{j}g_{k,j}^{l}a_{l,j}^{\dag }
\end{equation}%
which leads to%
\begin{equation}
H_{l}=\sum_{k}\varepsilon _{l,k}\eta _{l,k}^{\dag }\eta _{l,k},
\end{equation}%
where $\varepsilon _{l,k}$ is the corresponding eigenvalue of $H_{l}$\ for
the eigenfunction $g_{k,j}^{l}$. Site $j$ is defined as the wave node for
the eigen mode $k$ of graph $l$ if we have $g_{k,j}^{l}=0$. We denote the
wave node as $j(l,k)$, which reflects the property of the eigen state $\eta
_{l,k}^{\dag }\left\vert 0\right\rangle $ of $H_{l}$%
\begin{equation}
a_{l,j}\eta _{l,k}^{\dag }\left\vert 0\right\rangle =0,
\end{equation}%
where $\left\vert 0\right\rangle $\ is the vacuum state. Now we consider the
case of that \textit{all} the joint sites of the subgraph $l$ are the wave
nodes of eigen mode $k$. Under this condition, we have
\begin{equation}
H\left( \eta _{l,k}^{\dag }\left\vert 0\right\rangle \right) =H_{l}\left(
\eta _{l,k}^{\dag }\left\vert 0\right\rangle \right) =\varepsilon
_{l,k}\left( \eta _{l,k}^{\dag }\left\vert 0\right\rangle \right) ,
\end{equation}%
i.e., the eigen state $\eta _{l,k}^{\dag }\left\vert 0\right\rangle $ is
also the eigen state of the whole graph $H$. Then such a state represents
the trapping or bound state of a particle within the subgraph $l$ with
infinite life time. This rigorous conclusion has important implications in
the design of quantum network to store particles in the target region at
will. Figure \ref{fig1} represents an arbitrary graph of tight-binding
network within part of which particles can be confined without any leakage.
The whole graph can be decomposed into two subgraphs $l$ and $l^{\prime }$
which are connected via the couplings between the joint sites $(i,j,h)$\ and
$(i^{\prime },j^{\prime },h^{\prime })$. The perfect bound state can be
formed in subgraph $l$ as the eigen function of $H_{l}$ when it has wave
nodes on all the joint sites $(i,j,h)$. The existence of additional wave
nodes indicates the multiple bound states can be formed.
\begin{figure}[tbp]
\includegraphics[ bb=60 522 535 782, width=4.2 cm, clip]{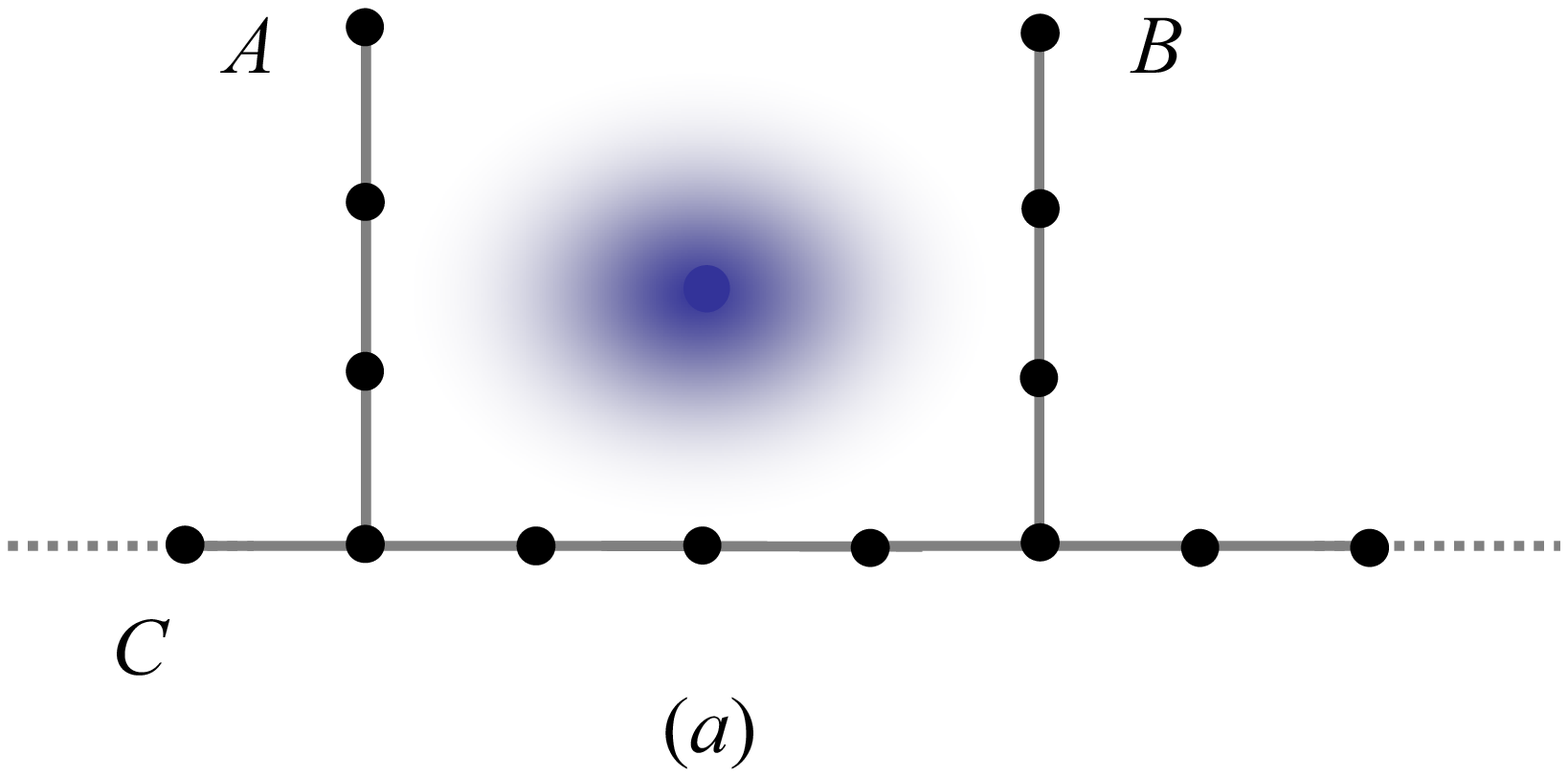}%
\includegraphics[ bb=60 521 535 781, width=4.2 cm, clip]{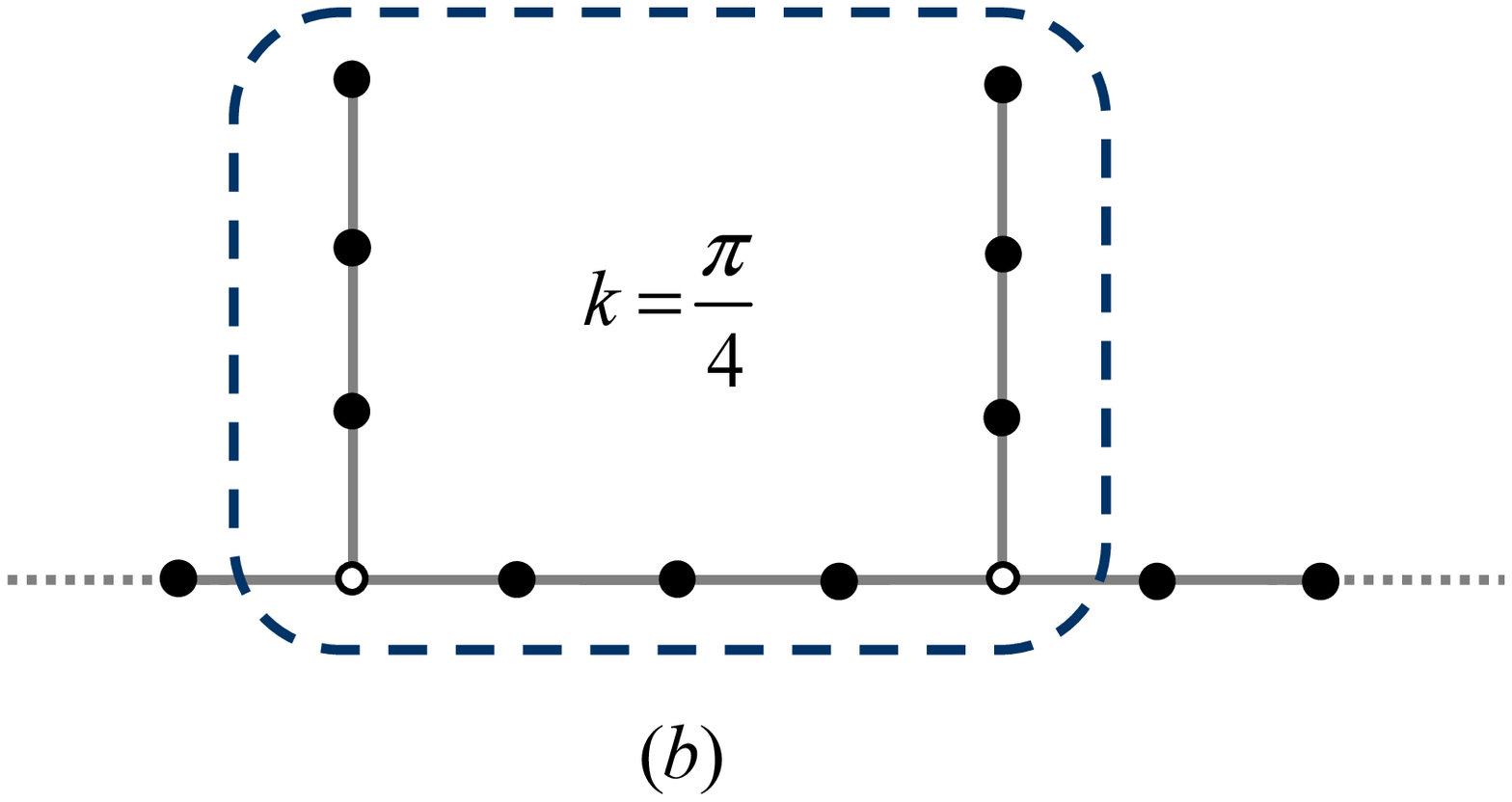} %
\includegraphics[ bb=60 521 535 781, width=4.2 cm, clip]{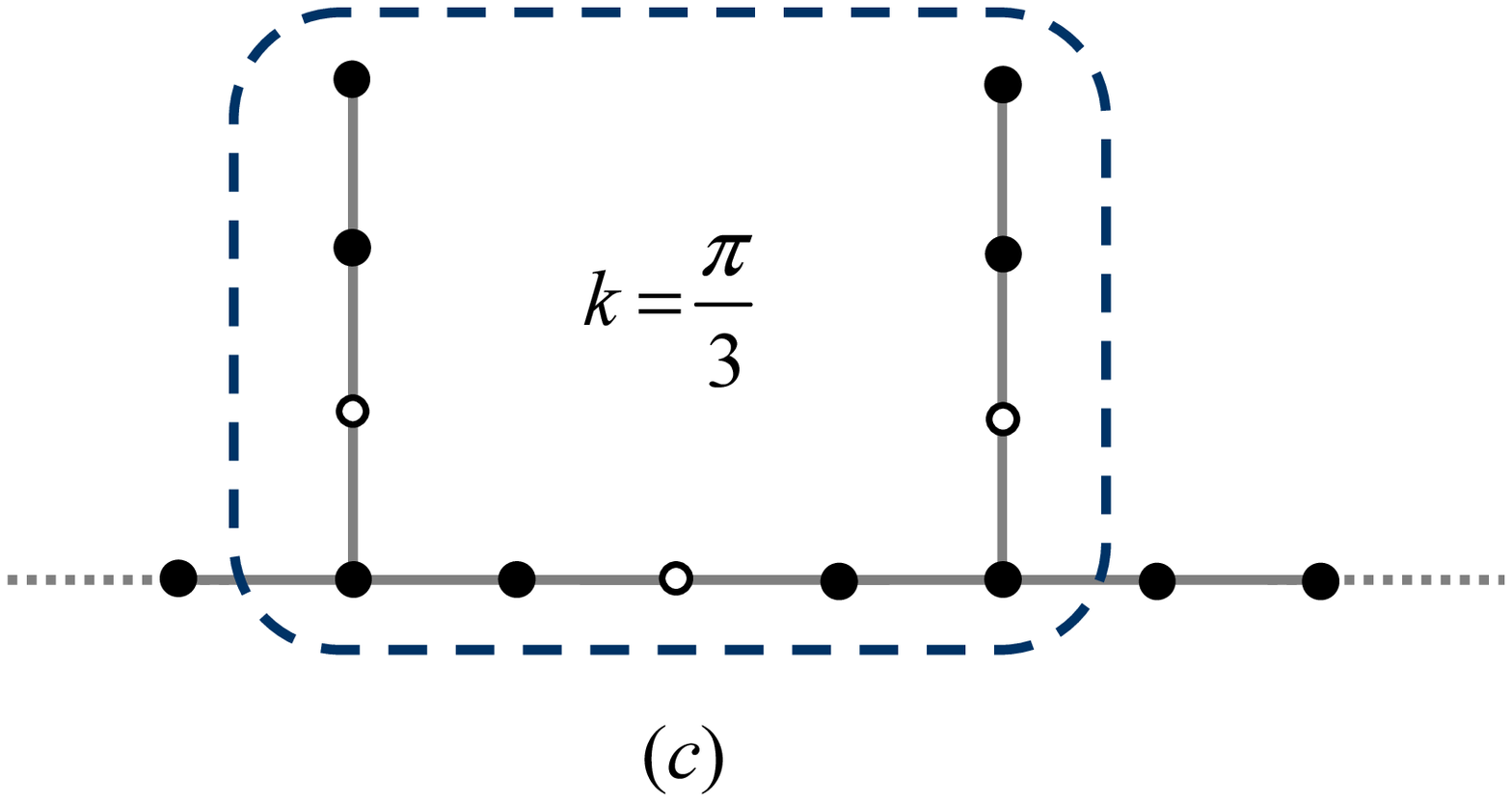}%
\includegraphics[ bb=60 521 535 781, width=4.2 cm, clip]{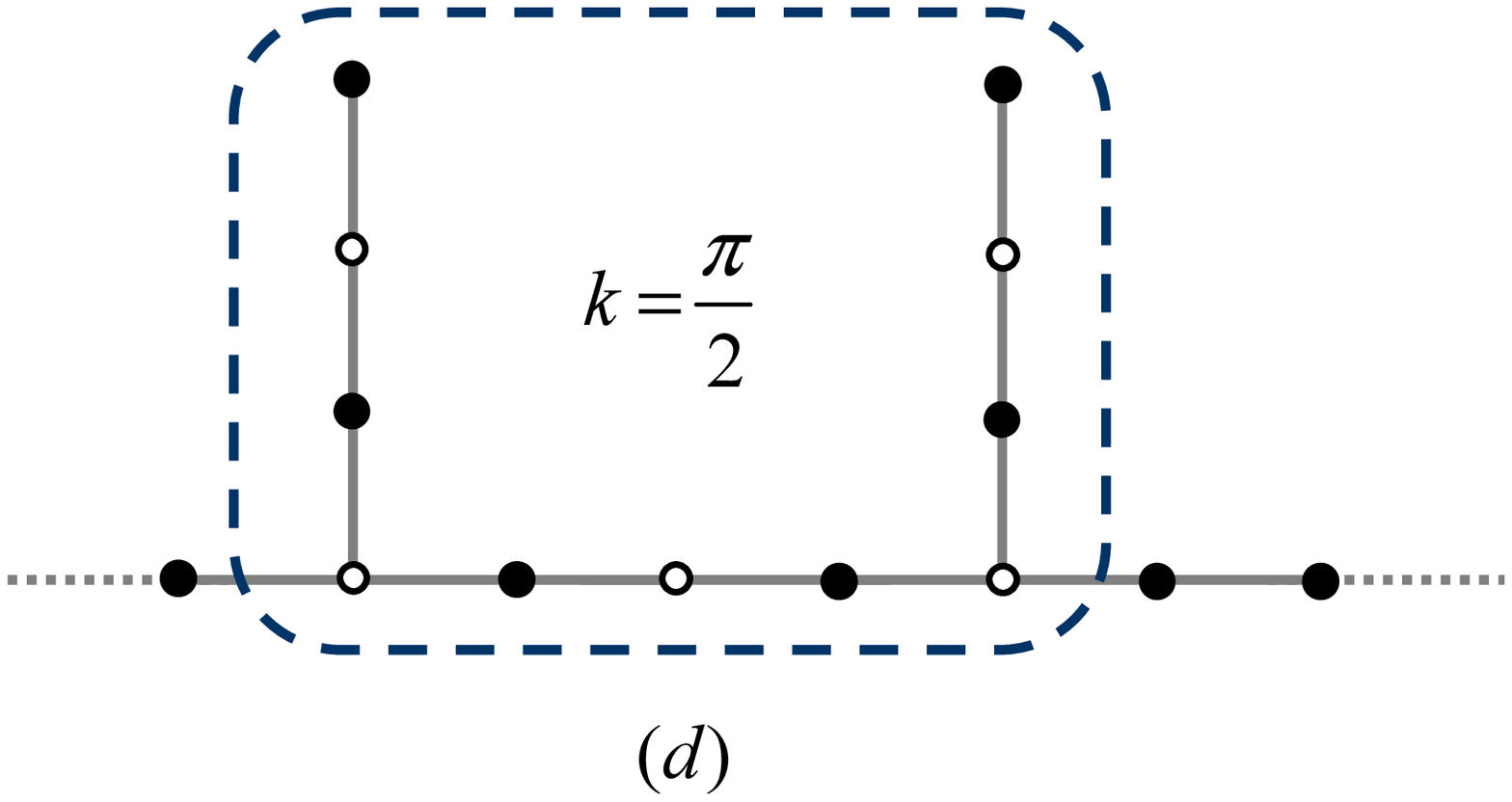}
\caption{(Color online) Configuration of $\protect\pi $-shaped lattice that
admits the formation of trapping particles. It consists of three chains $A$,
$B$ and $C$. Two three-site chains $A$ and $B$ are side coupled to the chain
$C$ with arbitrary number of sites (a). This graph can be decomposed into
three subgraphs, one of which is eleven-site chain enclosed by the dashed
rectangular. The single-particle eigen states of the eleven-site chain with
quasi momenta $k=\protect\pi /4,\protect\pi /3,$ and $\protect\pi /2$ have
two, three and four wave nodes, denoted by empty circles in (b), (c) and
(d). For states with $k=\protect\pi /4$ and $\protect\pi /2$, the joint
sites being all wave nodes, the particle can be trapped in the eleven-site\
chain, while state with $k=\protect\pi /3$ is not a bound state. States with
other values of $k$ can be analyzed accordingly.}
\label{fig2}
\end{figure}

\section{Demonstration configurations}

Now we investigate a class of practical examples to demonstrate the
application of the result above. We consider a system of $\pi $-shaped
lattice (Figure \ref{fig2}), consisting of an infinite chain side coupling
to two finite chains of length $N_{0}$ at the joint sites $1$ and $L $,
which has the Hamiltonian%
\begin{eqnarray}
H &=&H_{a}+H_{b}+H_{c}+H_{\text{joint}},  \label{H_abc} \\
H_{a}+H_{b} &=&-\kappa _{0}\sum_{i=1}^{N_{0}}(a_{i}^{\dag
}a_{i+1}+b_{i}^{\dag }b_{i+1}+\text{H.c.}),  \notag \\
H_{c} &=&-\kappa \sum_{i=-\infty }^{\infty }(c_{i}^{\dag }c_{i+1}+\text{H.c.}%
),  \notag \\
H_{\text{joint}} &=&-\kappa _{0}(a_{1}^{\dag }c_{1}+b_{1}^{\dag }c_{L}+\text{%
H.c.}),  \notag
\end{eqnarray}

where $a_{j}^{\dag }$ ($b_{j}^{\dag }$ and $c_{j}^{\dag }$) is the boson or
fermion creation operator at the $j$th site in the chain $a$ ($b$ and $c$).
The side coupling model was employed to depict coupled-cavity system for
stopping and storing light coherently \cite{Fan}. For the simple case with
the shortest side chains, i.e. $N_{0}=1$, the configuration is equivalent to
the atom-cavity system with single excitation \cite{ZhouL1, ZhouL2}, where
the side-site state represents the excited state of the two-level atom.

\begin{figure}[tbp]
\includegraphics[ bb=19 174 561 593, width=6.2 cm, clip]{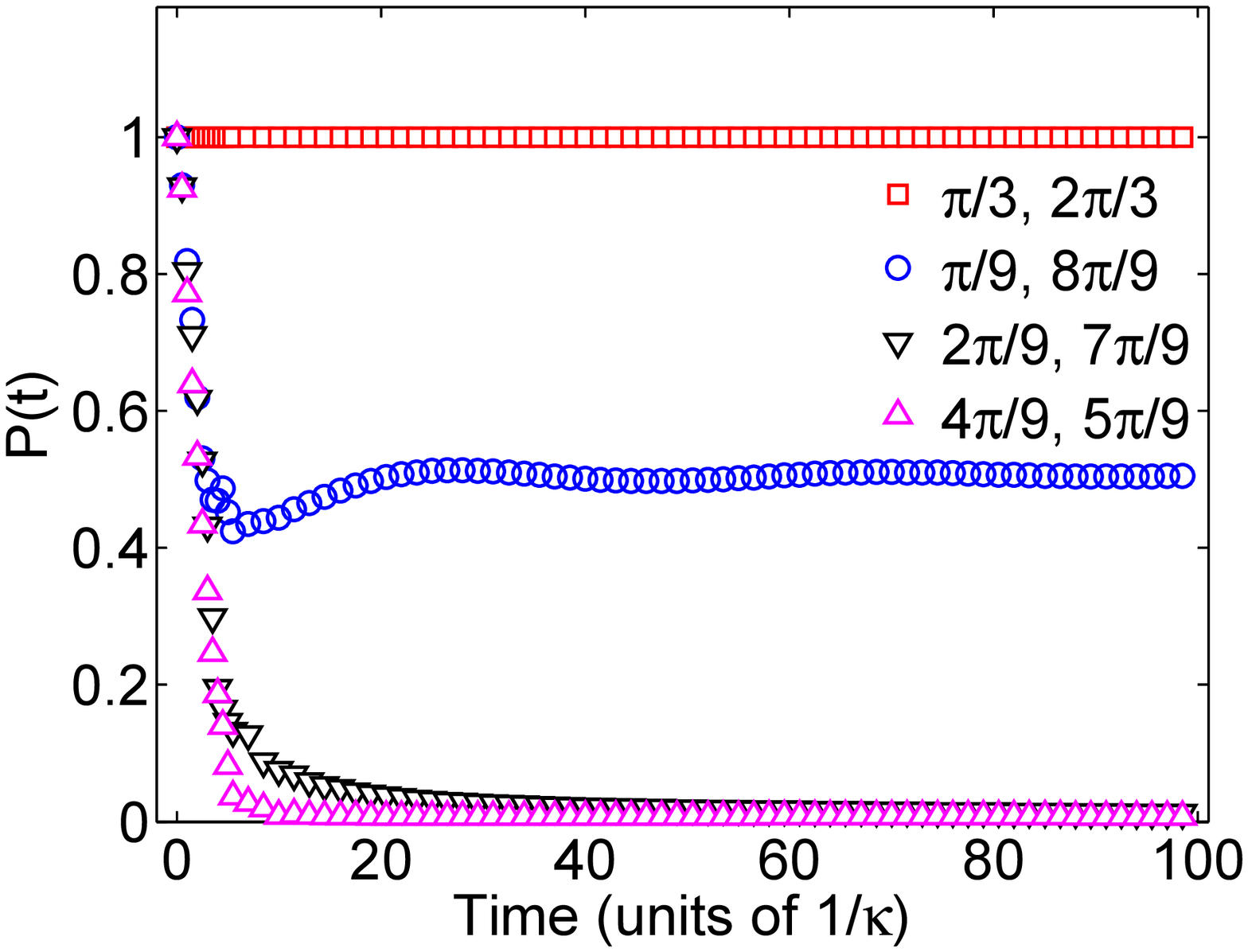} %
\includegraphics[ bb=19 174 561 593, width=6.2 cm, clip]{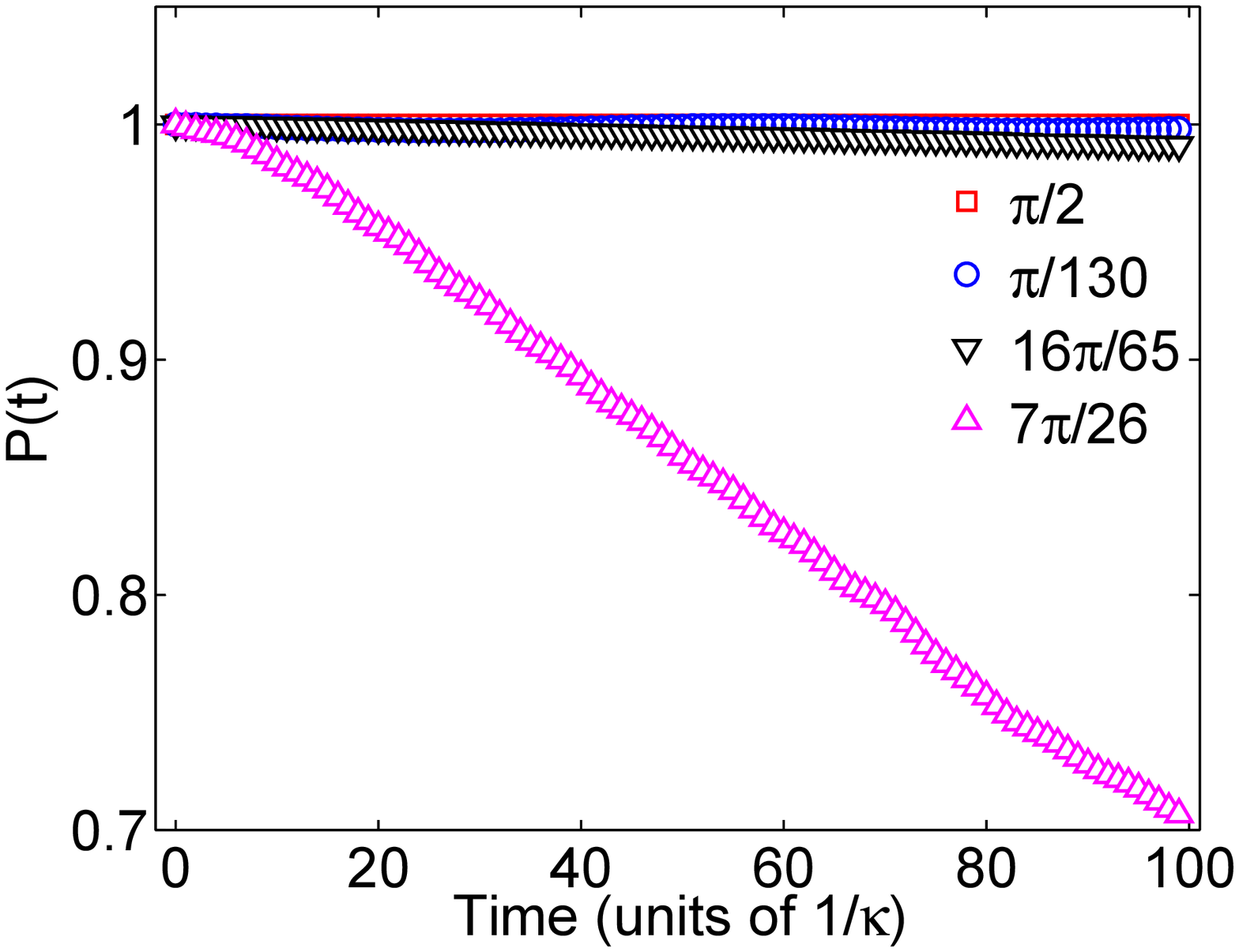}
\caption{(Color online)\ The probability $P(k,t)$ of a particle, initially
located in the state $\left\vert k\right\rangle $,$\ $remaining in the
subgraph. The simulations are performed in two typical systems with $N_{0}=2$%
, $L=4$ (upper panel) and $N_{0}=3$, $L=123$ (lower panel), respectively.
The shapes of all the curves can be classified into three types as mentioned
in the text.}
\label{fig3}
\end{figure}

First of all, we consider a simplest case: the hopping integrals are
identical for all chains, i.e., $\kappa =\kappa _{0}$. This graph can be
decomposed into three subgraphs: left chain, right chain and central chain
of $\Lambda =2N_{0}+L$ sites. The eigen wave functions of the central chain
are given by
\begin{equation}
g_{k,j}=\sqrt{\frac{2}{\Lambda +1}}\sin kj,\text{ }j\in \lbrack 1,\Lambda ],
\label{eigenstate}
\end{equation}%
where $k=n\pi /\left( \Lambda +1\right) $, $n\in \lbrack 1,\Lambda ]$, with
corresponding eigen values $-2\kappa \cos k$. These states possess the wave
nodes at
\begin{equation}
j_{k}=\frac{\left( \Lambda +1\right) m}{n},  \label{jk}
\end{equation}%
where $m$ are certain integers which ensure the existence of integer $j_{k}$
for a given $n$. Then in the case of that $\left\{ j_{k}\right\} $ cover the
joint sites $N_{0}+1$ and $N_{0}+L$ simultaneously, the corresponding eigen
states are the trapping states, i.e., a particle can be hold along the
central chain forever. An example of $N_{0}=3$, $L=5$ is depicted in Figure %
\ref{fig2}, where only typical cases with $k=\pi /4,\pi /3,$ and $\pi /2$
are presented. Actually, [Eq. (\ref{jk})] shows that states with $k=\pi /4,$
$\pi /2,$ and $3\pi /4$ have wave nodes at the joint sites. Therefore there
are three resonant bound states for this configuration. It has been proposed
that such kind of trapping state can act as a cavity when a boson system is
considered \cite{ZhouL2}. Remarkably, two peculiar features are identified.
First, the bound state has infinite life time in the ideal case without
decoherence since it is based on the mechanism of Fano interference rather
than two potential barriers. Second, the number of the cavity mode does not
solely depend on the size of the cavity $L$ like the case of using infinite
potential well for particle trapping. For example, taking $N_{0}=1$, one can
achieve a single mode cavity with $k=\pi /2$ for arbitrary odd $L$, but none
for even $L$. Meanwhile, it will be shown later that there is another type
of bound state, \textit{evanescent} bound state. Besides these exact bound
states, there exist eigen states of the subgraph which have nonzero, but
very small probability at the joint sites in the case of large $L$. Such
kind of state has finite but long life times, which is called quasi resonant
bound states. To demonstrate these concepts, we present a numerical
simulation of the damping process for various modes in two typical systems
with $N_{0}=2$, $L=4$ and $N_{0}=3$, $L=123$, respectively. A particle is
initially located in the subgraph in the eigen states $\left\vert
k\right\rangle $ [Eq. (\ref{eigenstate})]. We investigate the dynamics of
the states by computing the quantity
\begin{equation}
P(k,t)=\left\langle \sum_{i=1}^{N_{0}}(a_{i}^{\dag }a_{i}+b_{i}^{\dag
}b_{i})+\sum_{i=1}^{L}c_{i}^{\dag }c_{i}\right\rangle _{k,t},
\end{equation}%
where $\left\langle ...\right\rangle _{k,t}$ denotes the expectation value
of the probability of the particle within the subgraph for an evolved state $%
\exp \left( -iHt\right) \left\vert k\right\rangle $. Figure \ref{fig3} shows
the numerical simulation of $P(k,t)$\ as functions of the mode $k$ and time $%
t$ for a short $L$ in the upper plot while for a longer $L$ in the lower
plot. There are three types of curves in the two plots: (i) remaining
unitary; (ii) damping slowly; (iii) dropping drastically and then keeping at
a finite value. Cases (i) occurs in both two configurations, corresponding
to perfect resonant bound states. Case (ii) occurs in large-$L$ system,
corresponding to quasi resonant bound state (we omit such kind of curve in
the lower panel). Case (iii) occurs in small-$L$ system, corresponding to
another type of bound state, evanescent bound state, which will be discussed
in detail later.

A resonant bound-state configuration can be understood from the point of
view of interference. The bound state we constructed in this manner is the
standing-wave like state in the subgraph. In general, the formation of a
standing wave in a quantum system is due to the infinite potential barriers
which reflect the wave with \textit{any} momentum. Then there is no
additional condition for the distance between two barriers. In a
tight-binding network, a side coupled chain can act as the infinite
potential barriers for the incident wave with \textit{certain} momentum. As
an example, it can be readily shown by the method below that, for an
incident wave with $k=\pi /2$, the transmission coefficient $T$ through one
side coupled chain of length $N_{0}$ can be expressed as $T=\left[
1+(-1)^{N_{0}}\right] /2$. It has been discussed in Ref. \cite{ZhouL1,
ZhouL2} for the case of $N_{0}=1$. Besides the mirror condition $T=0$, a
matching distance between two side coupled chains is also required to form a
standing wave. This will be discussed below in the aid of exact results.

In the above analysis, the trapping subgraph is the simplest lattice, an
open chain. There are some a little more complicated subgraphs, the
hierarchical lattices, as the demonstration configurations. It has been
shown that \cite{ZLin,Cha96,Cha01} there are eigen wave functions of these
hierarchical lattices, whose amplitudes are zero at certain sites. When
these lattices are embedded in a network by linking the nodes only, the
trapping states are formed. Considering an arbitrary generation Vicsek
fractal as an example, there is eigen wave function whose amplitude is zero
at the center of every five-site cell. Then when such a lattice is embedded
in a network by linking the center sites only, the corresponding eigenstate
is the trapping state with respect to the network. Nevertheless, for the
hierarchical lattice itself, this eigenstate becomes an extended state as
its size grows up.

\section{Bethe Ansatz results}

We now turn to discuss the complete bound states in a subgraph by taking the
network of [Eq. (\ref{H_abc})] as an example. It is worthy to point out that
the bound states constructed by the before-mentioned method are not
complete. In the following it will be shown that there are two types of
bound states: \textit{resonant} and \textit{evanescent}. The former
describes trapped particle in a specified spatial region and the later
describes particle with an exponentially decaying probability beyond a
specified spatial region. In the following, we investigate this problem
based on the Bethe Ansatz approach. Actually, the bound-state wave functions
$\psi (j)$ of the Hamiltonian [Eq. (\ref{H_abc})] can be expressed as a
piecewise function over all sites%
\begin{eqnarray*}
\psi _{c}(j) &=&\left\{
\begin{array}{ll}
C_{1}e^{-ik\left( j-1\right) } & \text{ \ \ for }j\leq 1, \\
C_{2}e^{ikj}+C_{3}e^{-ikj} & \text{ \ \ for }2\prec j\prec L, \\
C_{4}e^{ik\left( j-L\right) } & \text{ \ \ for }j\geq L,%
\end{array}%
\right. \\
\psi _{a}(j) &=&A_{1}e^{iqj}+A_{2}e^{-iqj}\text{ \ \ for }1\leq j\leq N_{0},
\\
\psi _{b}(j) &=&B_{1}e^{iqj}+B_{2}e^{-iqj}\text{ \ \ for }1\leq j\leq N_{0}.
\end{eqnarray*}%
Here $\psi _{a,b,c}$ denote wave functions along chains $a$, $b$, and $c$,
respectively. The coefficients and momenta $C_{1,2,3,4}$, $A_{1,2}$, $%
B_{1,2} $, $k$,\ and $q$\ are determined by matching conditions and the
corresponding Schrodinger equations \cite{Hirsch}%
\begin{align}
\psi \left( j+0^{+}\right) =\psi \left( j+0^{-}\right) ,&  \label{continuity}
\\
-\kappa _{j+1,j}\psi \left( j+1\right) -\kappa _{j-1,j}\psi \left(
j-1\right) & =E\psi \left( j\right) .  \label{schrodinger}
\end{align}%
where $E$ is eigen energy, $\kappa _{j\pm 1,j}$ are the corresponding
hopping integrals. The solutions can be classified in two categories:
\textit{resonant} and \textit{evanescent} ones, which correspond to zero and
nonzero $C_{1,4}$, respectively.

For the resonant bound states, zero $C_{1,4}$ lead to zero particle
probability at the joint points, which is consistent with the above
mentioned rigorous results. In addition, the momenta $k$ and $q$ are
determined by equations
\begin{eqnarray}
\sin \left[ k\left( L-1\right) \right] &=&\sin \left[ q\left( N_{0}+1\right) %
\right] =0,  \label{resonant 1} \\
E &=&-2\kappa _{0}\cos q=-2\kappa \cos k.  \label{resonant 2}
\end{eqnarray}%
For simplicity, only simple cases with $\kappa _{0}=\kappa $ are considered
to demonstrate and explore the obtained rigorous results. The existence of
the solution requires $\left( L-1\right) m=\left( N_{0}+1\right) n$, where $%
n\in \lbrack 1,L-2]$ and $m\in \lbrack 1,N_{0}]$. Obviously, the resonant
bound states in the above mentioned example with $N_{0}=3$ and $L=5$ is the
simplest case of $m=n=1$, $2$, and $3$, corresponding to momenta $\pi /4$, $%
\pi /2$, and $3\pi /4$, respectively.

For the evanescent bound state, which possesses nonzero particle probability
at and around the joint points, the momenta $k$ and $q$ are determined by
equations%
\begin{eqnarray}
\frac{\kappa \zeta \left( k\right) }{\zeta \left( k\left( L-1\right) \right)
}\left[ e^{-ik\left( L-1\right) }\pm 1\right] &=&\frac{\kappa _{0}\zeta
\left( qN_{0}\right) }{\zeta \left( q\left( N_{0}+1\right) \right) },
\label{evanescent 1} \\
E=-2\kappa _{0}\eta \left( q\right) =-2\kappa \eta \left( k\right) , &&
\label{evanescent 2}
\end{eqnarray}%
where $\zeta \left( \theta \right) =$ $\left( e^{i\theta }-e^{-i\theta
}\right) /2$ and $\eta \left( \theta \right) =$ $\left( e^{i\theta
}+e^{-i\theta }\right) /2$. Taking $\kappa _{0}=\kappa $, $N_{0}=3$, and $%
L=5 $\ as an example, we have $q=k=0.382i$, $\pi +0.382i$, or $0.191i$, $\pi
+0.191i$, which correspond to symmetric and antisymmetric evanescent bound
eigen functions, respectively. Furthermore, for the case of $\kappa
_{0}=\kappa $, $N_{0}=2$, and $L=4$, plotted in Figure \ref{fig3}, we have $%
q=k=0.382i$ or $\pi +0.382i$. Accordingly, three initial states with momenta
$\pi /9$, $2\pi /9$, and $4\pi /9$, as well as their counterparts have
nonzero overlaps with the two evanescent bound states. We are then able to
obtain the long-time behavior of $P(t)$ as $0.5032$, $0.0027$, and $0.0058 $%
, which are in agreement with the plots in the upper panel of Figure \ref%
{fig3}.

\section{Scattering problems}

In general, trapping and scattering are two contrary phenomena which always
refer to localized and extended states. In the context of this paper, the
resonant bound state is essentially standing wave like, consisting of two
constituents: incident and reflected waves. On the other hand, the rigorous
result for such bound states has no restriction to the size and geometry of
the subgraph and is applicable to the scattering problem. This is another
main issue we want to stress in this paper.

For scattering problem, the input, output waveguides and the center system
should be involved. One can take the input waveguide, which is usually
semi-infinite chain, together with a part of the center system as the
subgraph. The resonant bound state in such a subgraph corresponds to a total
reflection. Actually, the trapping wave function within the input waveguide
region is the superposition of two opposite travelling plane waves with the
identical amplitudes. They correspond to the incident and total reflected
waves. And the eigen energy $E$\ of this trapping state is exactly the
transmission zero, i.e., $T(E)=0$. Taking the above $\pi $-shaped lattice as
an illustrated example, the subgraph containing the input waveguide is
depicted by the Hamiltonian%
\begin{equation}
H_{in}=-\kappa \left( \sum_{i=1}^{N_{0}}a_{i}^{\dag }a_{i+1}+a_{1}^{\dag
}c_{1}+\sum_{i=-\infty }^{1}c_{i}^{\dag }c_{i+1}+\text{H.c.}\right) ,
\label{H_in}
\end{equation}%
which is a uniform semi-infinite chain. The resonant bound states must have
a node at site $c_{1}$\ with energy $E=-2\kappa \cos q$, where $q$\ is
determined by the position of the node, $\sin \left[ q\left( N_{0}+1\right) %
\right] =0$.

Now we consider the scattering problem of the $\pi $-shaped lattice,
demonstrating the relation linking the scattering state and resonant bound
state in the framework of the Bethe Ansatz. It is worth to note that many
efforts have been devoted to discuss critically the effect of a dangling
side coupled chain on the spectrum and transmission properties of a linear
chain, including the Fano resonance, by approximate approaches \cite%
{AY,PF,Cha06,Cha07}.

In a $\pi $-shaped lattice, the scattering wave function has the form\textbf{%
\ }%
\begin{eqnarray*}
\psi _{c}(j) &=&\left\{
\begin{array}{ll}
e^{ik\left( j-1\right) }+re^{-ik\left( j-1\right) } & \text{ \ \ for }j\leq
1, \\
Ae^{ik\left( j-1\right) }+Be^{-ik\left( j-1\right) } & \text{ \ \ for }%
2\prec j\prec L, \\
te^{ik\left( j-1\right) } & \text{ \ \ for }j\geq L,%
\end{array}%
\right. \\
\psi _{a}(j) &=&C_{a}e^{iqj}+D_{a}e^{-iqj},\text{ \ \ for }1\leq j\leq N_{0},
\\
\psi _{b}(j) &=&C_{b}e^{iqj}+D_{b}e^{-iqj},\text{ \ \ for }1\leq j\leq N_{0}.
\end{eqnarray*}%
where $r$\ and $t$\ are reflection and transmission amplitudes for an
incident wave with momentum $k$. Similarly, applying the matching conditions
[Eq. (\ref{continuity})] and the corresponding Schrodinger equations [Eq. (%
\ref{schrodinger})], we then obtain\textbf{\ }%
\begin{equation}
t=\frac{\alpha ^{2}\sin ^{2}k}{\alpha ^{2}\sin ^{2}k-i\alpha \beta \sin
k+\left( \beta /2\right) ^{2}\left[ e^{i2k\left( L-1\right) }-1\right] }
\label{tran}
\end{equation}%
where $\alpha =\kappa \sin \left[ q\left( N_{0}+1\right) \right] $\ and $%
\beta =\kappa _{0}\sin \left( qN_{0}\right) $. Note that zero $\alpha $\
leads to vanishing of $t$, while zero $\beta $\ leads to vanishing of $r$.
The former and latter are in agreement with the conclusions of the above
analysis from the interference point of view for the total reflection and
resonant transmission, respectively.

From Eq. (\ref{tran}), the transmission probability has the form%
\begin{equation}
T=\frac{\alpha ^{4}\sin ^{4}k}{\alpha ^{4}\sin ^{4}k+\left( \beta /2\right)
^{2}\left( \beta ^{2}+4\alpha ^{2}\sin ^{2}k\right) \sin ^{2}\left[ k\left(
L-1\right) -\delta \right] }  \label{T}
\end{equation}%
where $\tan \delta =2\alpha \sin k/\beta $. Eq. (\ref{T}) allows the
analytical investigation on the transmission features. First, it is found
that the total reflection condition coincides with the resonant bound
condition Eq. (\ref{resonant 1}). It indicates the conclusion that an
incident wave is totally reflected by the side coupled chains if its energy
exactly equals to the resonant bound state energy. The same conclusion has
been obtained for some similar systems \cite{AY,PF,Cha06,Cha07}. This is a
direct result from the fact that the scattering of\ any dangling side
coupled chain is isotropic for the incident waves from both sides along the
waveguide. On the other hand, the resonant transmission condition is also
easy to be understood from the aspect of wave nodes in the subgraph. In fact
equation $\sin \left( qN_{0}\right) =0$ indicates the effective
disconnection of the wave guide from the side coupled system.

Second, for a\ fixed $N_{0}$, the common transmission zeros and reflection
zeros for arbitrary $L$\ can be simply determined by $\alpha =0$\ and $\beta
=0$, respectively. More precisely, for the incident waves with $k_{\min
}=\cos ^{-1}\left\{ \left( \kappa _{0}/\kappa \right) \cos \left[ n\pi
/\left( N_{0}+1\right) \right] \right\} $, $n\in Z$, we have $T=0$, while
the one with $k_{\max }=\cos ^{-1}\left[ \left( \kappa _{0}/\kappa \right)
\cos \left( n\pi /N_{0}\right) \right] $, we have $T=1$. The rest reflection
zeros are $L$-dependent and determined by $\sin ^{2}\left[ k\left(
L-1\right) -\delta \right] =0$.

\begin{figure}[tbp]
\includegraphics[ bb=47 184 390 602, width=4.2 cm, clip]{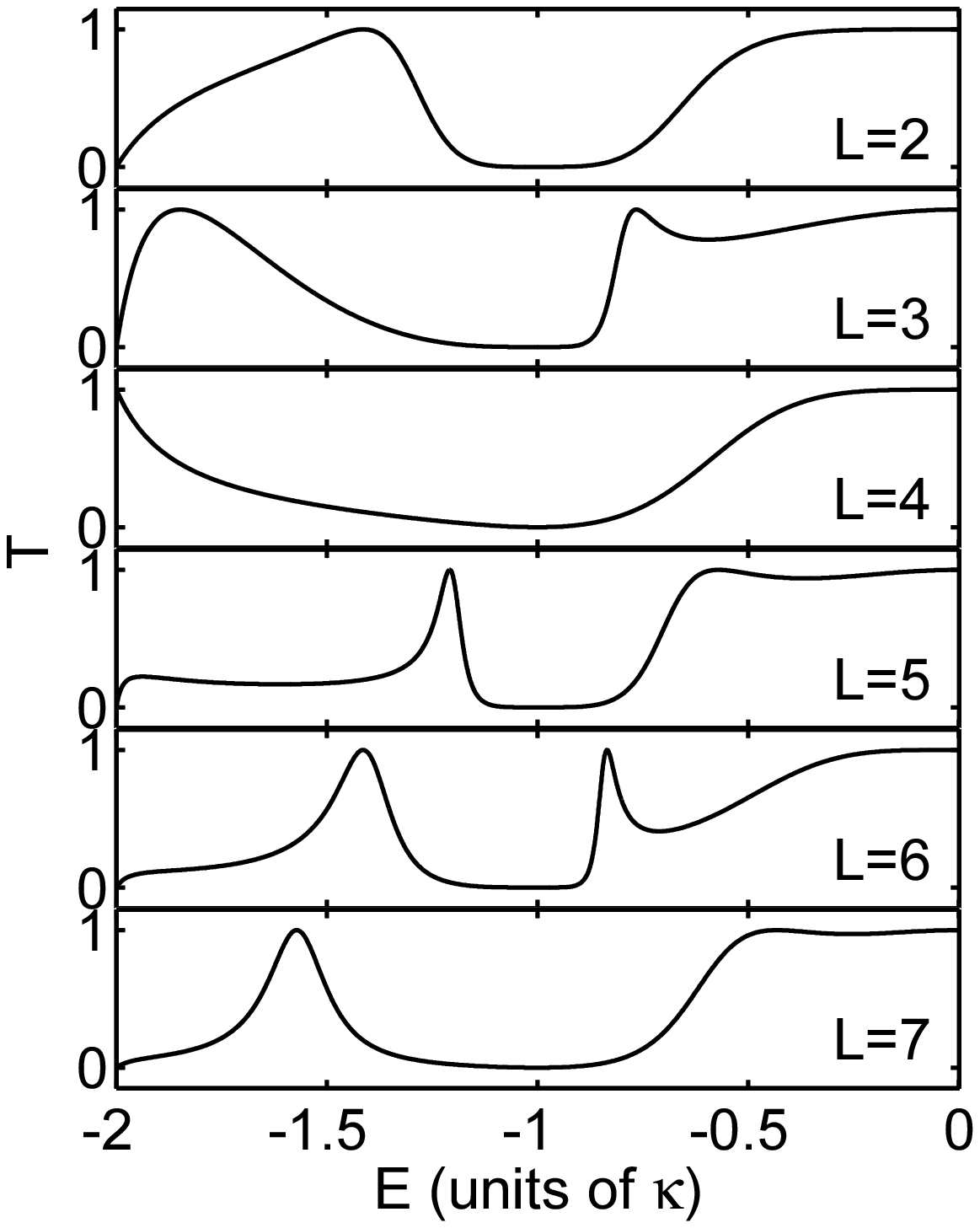} %
\includegraphics[ bb=47 184 390 602, width=4.2 cm, clip]{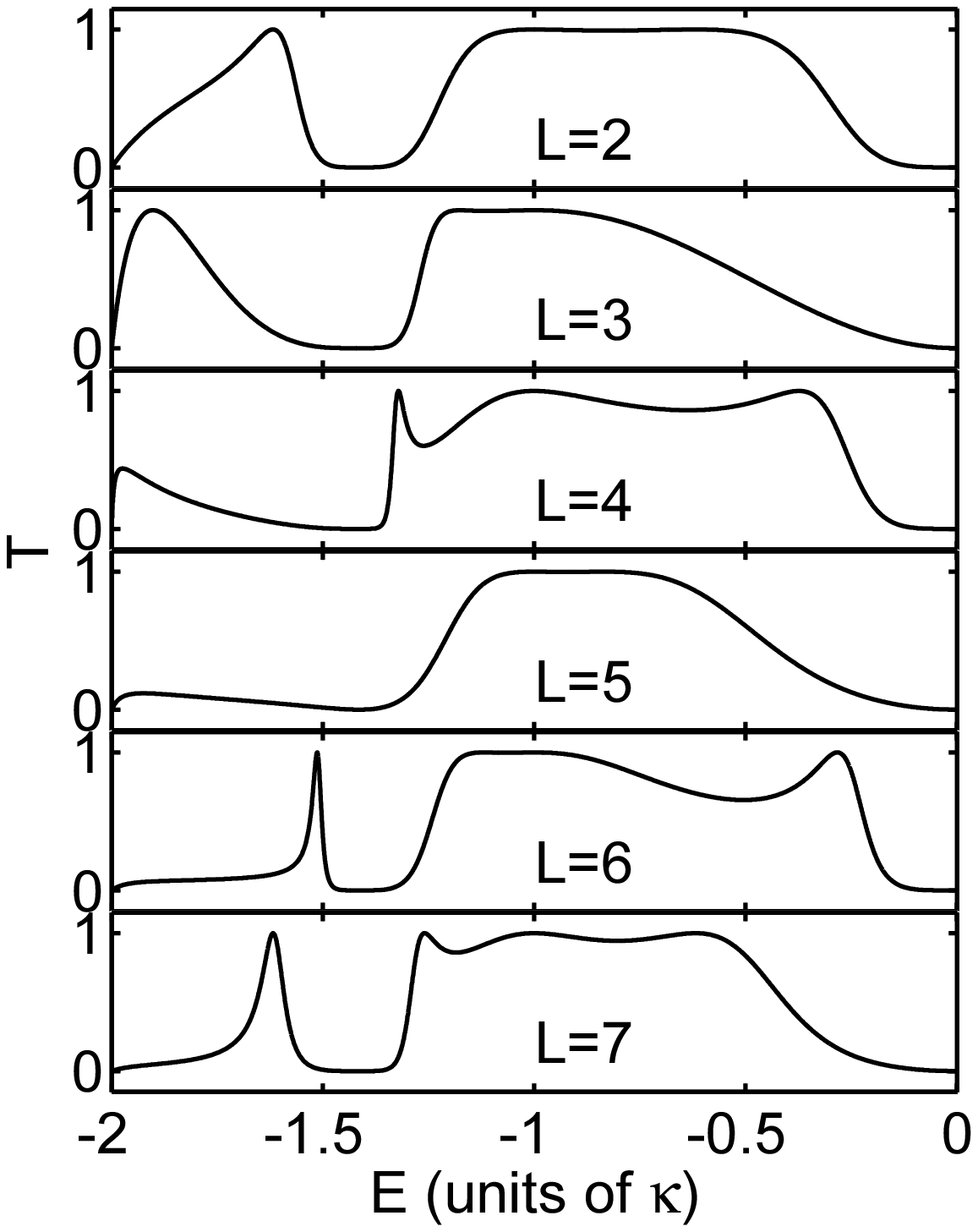}
\caption{The plots of $T(E)$ from Eq. (\protect\ref{T}) for the systems of $%
N_{0}=2$ (left), $3$ (right) with different $L$. }
\label{fig4}
\end{figure}

The transmission spectra are plotted for $\kappa =\kappa _{0}$, $N_{0}=2$\ ,$%
3$\ and different $L$\ in Fig. \ref{fig4} as illustration. We can see that
the common transmission zeros occur at $E=-1$\ for $N_{0}=2$; $E=-\sqrt{2}$\
for $N_{0}=3$, while the common reflection zeros occur at $E=0$\ for $%
N_{0}=2;$\ $E=-1$\ for $N_{0}=3$, which are in agreement with the above
analysis. From the plots, one can find that it does not exhibit perfect Fano
line shape. Nevertheless, the peak and dips profiles are the direct result
of interference result from subwaves in different paths. Actually, the
formations of $k_{\min }$\ and $k_{\max }$\ correspond to the complete
destructive and constructive interferences.

\begin{figure}[tbp]
\includegraphics[ bb=33 181 548 587, width=7 cm, clip]{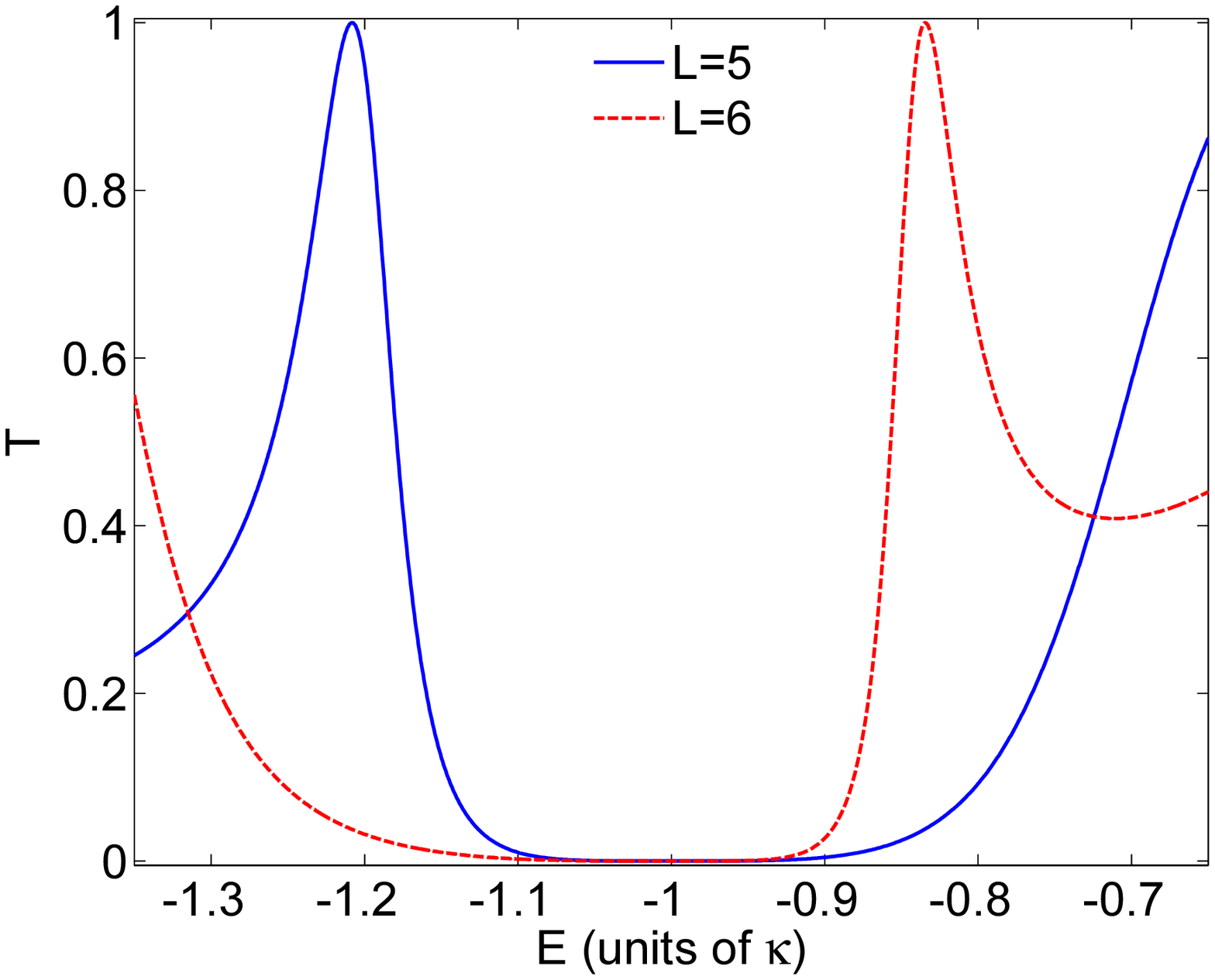} %
\includegraphics[ bb=33 181 548 587, width=7 cm, clip]{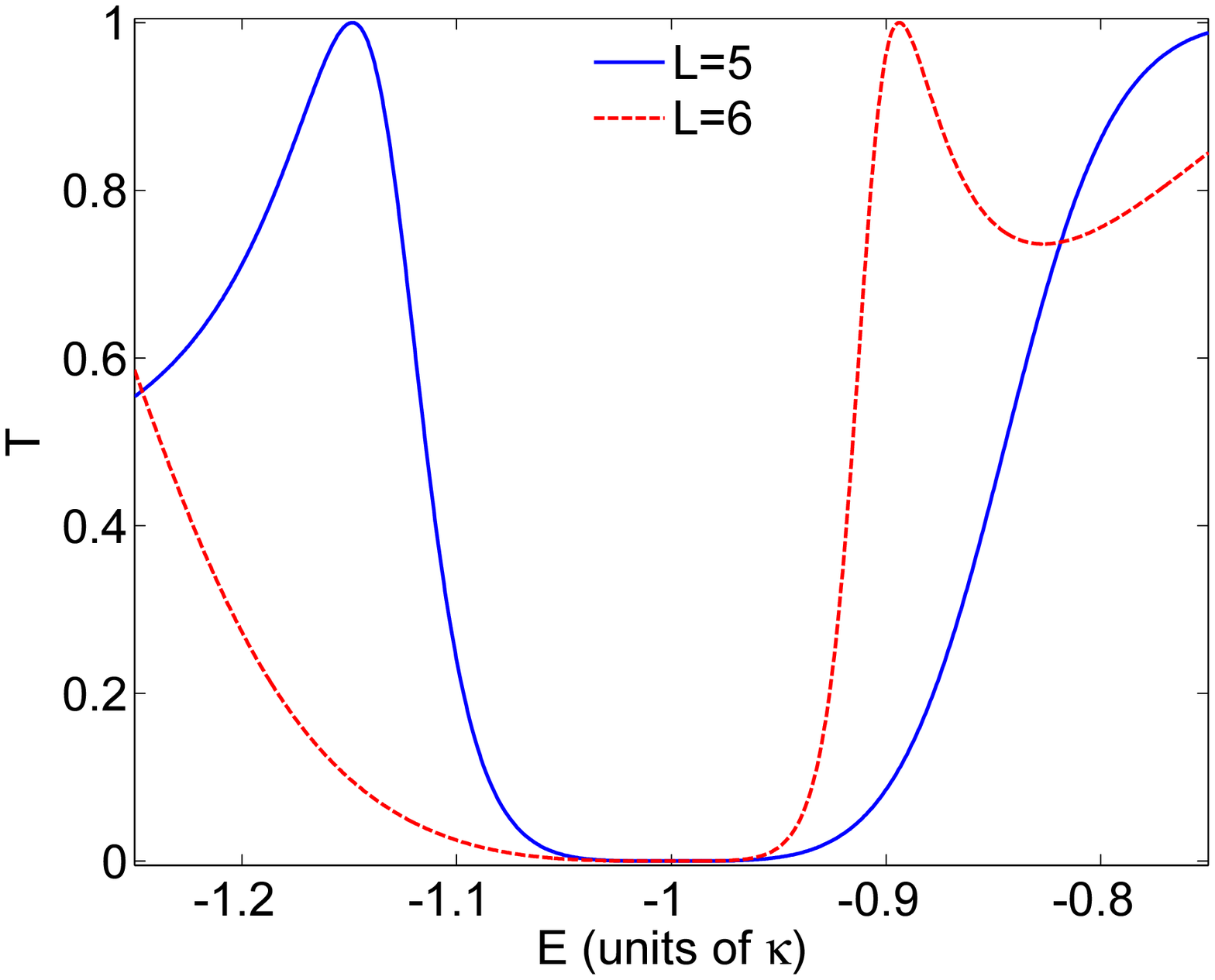}
\caption{(Color online) Transmission probability $T(E)$ for the
configurations with $N_{0}=2$ (up panel), $5$ (down panel) and $L=5$ (solid
blue line), $L=6$ (dashed red line). The plots show the evident swapping of
peak-dip profiles.}
\label{fig5}
\end{figure}

Now we focus on the $L$-dependent reflection zeros. Consider a system with
fixed $L=L_{0}$, the $L$-dependent reflection zeros occur at $k_{0}$, which
satisfies%
\begin{equation}
\sin ^{2}\left[ k_{0}\left( L_{0}-1\right) -\delta \right] =0.
\label{k_0 eq}
\end{equation}%
Meanwhile, for a system with $L=L_{0}+m$, the corresponding transmission
coefficient obeys%
\begin{equation}
T\left( k_{0},L_{0}+m\right) =T\left( k_{0},L_{0}-m\right)
\end{equation}%
for $L_{0}-m\succ 0$, due to the identity%
\begin{equation}
\sin ^{2}\left[ k_{0}\left( L_{0}+m-1\right) -\delta \right] =\sin
^{2}\left( mk_{0}\right) .
\end{equation}%
This fact leads to an interesting conclusion. For a certain $k_{0}$, if
there are two systems $L$\ and $L^{\prime }$\ that satisfy $T\left(
k_{0},L\right) =$\ $T\left( k_{0},L^{\prime }\right) =1$, there should exist
a series of different $L,L^{\prime },L^{\prime \prime },L^{\prime \prime
\prime },...,$\ satisfy $T\left( k_{0},L^{\prime \prime }\right) =$\ $%
T\left( k_{0},L^{\prime \prime \prime }\right) =...=1$. Especially, applying
this conclusion for $m=1$\ case, it follows that there is no $k_{0}$\ to
satisfy $T\left( k_{0},L\right) =$\ $T\left( k_{0},L+1\right) =1$, except
the common reflection zeros. In other words, there is no $L$-dependent
reflection zeros for $L$\ and $L+1$\ meeting at the same $k$. This feature
enhance the probability of the occurrence of so called peak-dip swapping as $%
L$\ changes \cite{Cha06,Cha07}.

For a fixed $N_{0}$, one can always find two systems with successive $L$,
that they have at least one peak (reflection zero) located at each side of a
common dip (transmission zero). Since there is only one peak at each $k_{0}$%
, the peak-dip swapping profile is formed in the vicinity of a common dip.
Here we exemplify this point by investigating the cases with $\kappa =\kappa
_{0}$\ and small $N_{0}$. For $N_{0}=2$,\ one of the common transmission
zero is $k_{\min }=\pi /3$,\ while the $L$-dependent reflection zeros are
determined by $\sin ^{2}\left[ k_{0}\left( L_{0}-1\right) -\delta \right] $\
$=0$. The closest (or closer) solution of $k_{0}$\ around $k_{\min }=\pi /3$%
\ are the left one $k_{0L}=0.29\pi $\ ($E=-1.21$) for $L=5$, and the right
one $k_{0R}=0.36\pi $\ ($E=-0.84$) for $L=6$. The profiles of the
corresponding transmission spectra are plotted in Fig. \ref{fig5} (upper),
which exhibit the same character as the one in Fig. 7 of Ref. \cite{Cha06}.
Another example for $N_{0}=5$\ and $L=5$, $6$, is also plotted in Fig. \ref%
{fig5} (lower). One can see the occurrence of the profile of evident
peak-dip swapping.

\section{Summary}

In summary, we show in this paper, within the context of a tight-binding
model, that a particle can be trapped in a nontrivial subgraph. As an
application, we examine concrete networks consisting of a $\pi $-shaped
lattice. Exact solutions for such types of configurations are obtained to
demonstrate and supplement the rigorous results. It is shown that there are
two types of bound states: resonant and evanescent. We also link the
trapping rigorous result to the scattering problem for such a subgraph being
embedded in a one-dimensional chain as the waveguide. It is shown that an
incident wave experiences total reflection under certain condition. Finally,
we also investigate the scattering features of the $\pi $-shaped lattice in
the framework of the Bethe Ansatz. Such rigorous results are expected to be
necessary and insightful for quantum control and engineering.

We acknowledge the support of the CNSF (Grants No. 10874091 and No.
2006CB921205).

\end{document}